\documentclass[prl,twocolumn,showpacs,amsmath,amssymb]{revtex4}

\usepackage{graphicx}

\begin{document}
\title{Transition between electron localisation and antilocalisation in graphene}
\author{F.~V.~Tikhonenko}
\author{A.~A.~Kozikov}
\author{A.~K.~Savchenko}
\author{R.~V.~Gorbachev}
\affiliation{School of Physics, University of Exeter, EX4 4QL, U.K.}

\pacs{73.23.-b, 72.15.Rn, 73.43.Qt}

\begin{abstract}
The wave nature of electrons in low-dimensional structures manifests itself in conventional electrical measurements as a quantum correction to the classical conductance. This correction comes from the interference of scattered electrons which results in electron localisation and therefore a decrease of the conductance. In graphene, where the charge carriers are chiral and have an additional (Berry) phase of $\pi$, the quantum interference is expected to lead to anti-localisation: an increase of the conductance accompanied by negative magnetoconductance (a decrease of conductance in magnetic field). Here we observe such negative magnetoconductance which is a direct consequence of the chirality of electrons in graphene.  We show that graphene is a unique two-dimensional material in that, depending on experimental conditions, it can demonstrate both localisation and anti-localisation effects.  We also show that quantum interference in graphene can survive at unusually high temperatures, up to $T\sim$200 K.
\end{abstract}

\maketitle

Quantum interference of electrons in disordered low-dimensional structures has been a subject of intensive studies \cite{Beenakker}. It can be seen in simple measurements of the conductance, as a correction to the classical (Drude) conductivity.  Classically, electrons in a conductor are treated as particles which are scattered by impurities, while quantum mechanically electrons will not only scatter but also interfere with each other. Quantum interference occurs on closed electron trajectories, Fig. 1a, where two electron waves travel in both directions and interfere at the point of intercept.   As the two paths are identical, the phase of the waves is the same and the interference constructive.  This increases the probability of the electron to stay in the intercept region (to be weakly localised) and decreases the overall electrical conductance. In experiment, the quantum correction to the conductance can be detected by applying magnetic fields perpendicular to the current: magnetic flux through the closed trajectory adds a phase difference to the two waves and destroys the interference.  Therefore, the measured conductance increases in magnetic field -- positive magnetoconductance (MC), $\Delta \sigma (B)=\sigma (B)-\sigma (B=0)>0$. Until now there has been one known case where the quantum correction to the conductance is \emph{positive} and accompanied by a \emph{negative} MC.  Such \emph{anti-localisation} of electrons occurs in materials with strong spin-orbit scattering \cite{AKS, Bergman, Meier} where scattered electrons flip their spins and the two waves interfere destructively.

In graphene electrons are very different from those in conventional two-dimensional (2D) systems. Importantly for interference, they are \emph{chiral}. This means that they possess an additional quantum number, the pseudo-spin, and have a Berry phase of $\pi$ -- the additional phase that the wavefunction will acquire if an electron completes a full-circle trajectory.  The direct manifestation of chirality is the anomalous propagation of electrons through potential barriers (the Klein paradox \cite{Klein,Kim}). The other consequence of the chiral nature and the Berry phase should be electron anti-localisation, without strong spin-orbit scattering which in graphene is very weak. Due to the Berry phase of $\pi$ the two trajectories in Fig. 1a will meet  in anti-phase: one will acquire an extra phase of $\pi/2$ and the other, moving in the opposite direction, the phase $-\pi/2$.

It was expected that anti-localisation in graphene could only be seen in very clean layers without defects in the crystal structure. Similar to other 2D systems the magnitude of the quantum correction in graphene depends on inelastic scattering of electrons characterised by the dephasing time $\tau_{\phi}$: a change of the electron energy breaks phase coherence and limits the size of the interfering trajectories in Fig. 1a. Unusually, quantum interference in graphene is also controlled by elastic (energy conserving) scattering caused by imperfections in the crystal structure. These elastic processes are described by two characteristic times:  $\tau_*$ and $\tau_i$. Graphene's band structure has two valleys, and quantum interference of electrons in one valley can be suppressed by defects with the size of the lattice spacing, as well as dislocations and ripples \cite{MorpurgoPRL06,McCannPRL06,Morozov}. (Such defects break the chirality, while dislocations and ripples produce an effective random magnetic field which destroys the interference.) The combined effect of this intra-valley scattering is characterised by time $\tau_*$. Inter-valley scattering by the sharp defects (such as the edges of the sample) that are able to change strongly the electron momentum is characterised by the time $\tau_i$. As the two valleys have opposite chirality, intervalley scattering has a restoring effect on the suppressed quantum interference, and so gives rise to electron localisation. In recent experiments both $\tau_*$ and $\tau_i$ were found to be small, and this resulted in a negative quantum correction in the studied graphene-based structures \cite{Wu,Gorbachev,Tikhonenko,Horsell}.
\begin{figure}[t]
\includegraphics[width=\columnwidth]{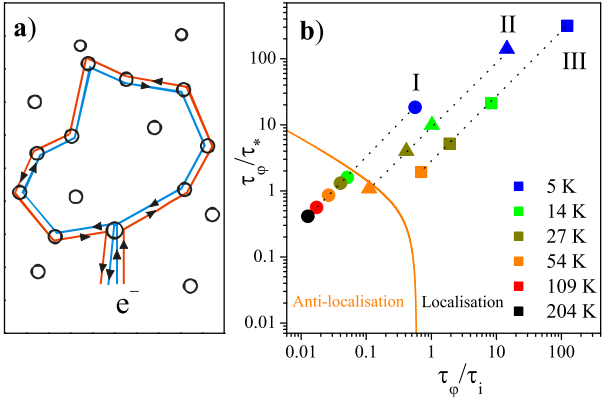}
\caption{(a) The trajectories of an electron scattered by impurities that give rise to quantum interference. (b) A diagram of scattering times related to quantum interference in graphene. The solid curve separates the regions of electron localisation and anti-localisation. Points are experimental values found from the analysis of the magnetoconductance using Eq. 1, for three regions of electron density. }\label{fig:one}
\end{figure}

In this work we present direct evidence of the Berry phase $\pi$ of electrons in graphene by observing negative low-field MC. We show that in mechanically exfoliated graphene one can detect electron anti-localisation even in the presence of the intra- and inter-valley scattering, provided the characteristic times $\tau_*$ and $\tau_i$ are large enough compared with the dephasing time $\tau_{\phi}$. We show that by increasing the temperature and decreasing the electron density one can achieve the conditions of anti-localisation, and that by changing the experimental conditions for the same graphene sample a transition between localisation and anti-localisation can be seen.

The theory \cite{McCannPRL06} of quantum interference predicts the following MC in graphene:
\begin{eqnarray}
\Delta\sigma(B)&=&\frac{e^{2}}{\pi h}\cdot\left(F\left(\frac{\tau_B^{-1}}
{\tau_{\phi}^{-1}}\right)-F\left(\frac{\tau_B^{-1}}{\tau_{\phi}^{-1}+
2\tau_{i}^{-1}}\right)\right.\nonumber\\
&&-2F\left.\left(\frac{\tau_B^{-1}}{\tau_{\phi}^{-1}+
\tau_{i}^{-1}+\tau_{*}^{-1}}\right)\right)\; .\label{eqn:one}
\end{eqnarray}
Here $F(z)=\ln{z}+\psi{\left(0.5 + z^{-1} \right)}$, $\psi(x)$ is the digamma function, $\tau_B^{-1}=4eDB/\hbar$ and $D$ is the diffusion coefficient. (The theory assumes that the momentum relaxation rate $\tau_p^{-1}$ is the highest in the system and comes from the Coulomb charges, and does not affect the electron interference.) Negative MC corresponding to anti-localisation is described by the second and third (negative) terms. In the absence of intra- and inter-valley scattering, $\tau_{i,*}\rightarrow \infty$,  $\Delta\sigma(B)$ is totally determined by the third term. This situation would correspond to anti-localisation in a defect-free graphene layer. In the opposite case of strong intra- and inter-valley scattering (small $\tau_*$ and $\tau_i$), the negative terms are suppressed and the first (positive) term dominates, which corresponds to electron localisation.  It is this situation that was realised in the recent experiments  \cite{Tikhonenko,Horsell} where the negative terms, originating from the chirality of electrons, were not large enough to change the sign of the low-field MC to negative.

Equation 1, however, demonstrates that anti-localisation can in principle be detected in real graphene samples with finite intra- and inter-valley scattering. Using the fact that the function $F(z)$ for $z\ll 1$ (at small magnetic fields) can be represented by a simple quadratic dependence $F(z)=z^2/24$, we simplify Eq. 1 for these fields as
\begin{eqnarray}
\Delta\sigma(B)&=&\frac{e^{2}}{24\pi h}\cdot\left(\frac{4eDB}{\hbar}\right)^{2}\left(1- \frac{1}
{(1+2\tau_{\phi}/\tau_i)^2}\right.\nonumber\\
&&\left. - \frac{2}{(1+\tau_{\phi}/\tau_i+\tau_{\phi}/\tau_{*})^2}
\right)\; .\label{eqn:two}
\end{eqnarray}
It is clear then that the negative sign of MC is determined by the sign of the expression in the brackets and can be reached for a range of the three characteristic times.  Fig. 1b shows a diagram with a curve $\Delta\sigma=0$ found from Eq. 2 that separates the regions of positive and negative MC (localisation and anti-localisation). The favorable conditions for the observation of negative MC correspond to small $\tau_{\phi}/\tau_*$ and $\tau_{\phi}/\tau_i$. In experiment, the ratio of the times can be changed by increasing temperature and carrier density.  Increasing the temperature decreases the dephasing time $\tau_{\phi}$, and increasing electron density will decrease $\tau_i $ due to increasing the density of electron states  - this trend was detected in Ref.\cite{Tikhonenko}.

The studied sample is produced by mechanical exfoliation of graphite and deposited on an oxidized Si wafer \cite{NovoselovScience04}. Using electron-beam lithography a six-terminal Hall bar is formed from the flake with a width of 2 $\mu$m and a length between the potential probes (Au/Cr) of 22.5 $\mu$m.  To improve the quality of the sample, it was annealed in vacuum at a temperature of 140$^{\circ}$C for two hours before cooling down in a cryostat. The inset shows a schematic representation of the measurement circuit. The upper inset shows a measurement of the Quantum Hall effect where plateaux at half-integer filling factors are evidence that the sample is monolayer graphene \cite{Monolayer}. The mobility of electrons outside the Dirac (electro-neutrality) region is $\sim$ 12000 cm$^2$ V$^{-1}$s$^{-1}$. The resistance $R$ as a function of the gate voltage $V_g$ shows a peak at the Dirac point where the density of charge carriers is zero, Fig. 2a. The bars indicate three regions of the gate voltage (each of the size $\Delta V_g=$1 V) where the magnetoconductance is measured. To average out the effect of universal conductance fluctuations \cite{Kechedzhi}, the resistance at each magnetic field, which is changed with a step from 1 to 10 mT, is averaged over each of the regions. These measurements are then repeated at each studied temperature in the range $T$=5 - 200 K.
\begin{figure}[t]
\includegraphics[width=\columnwidth]{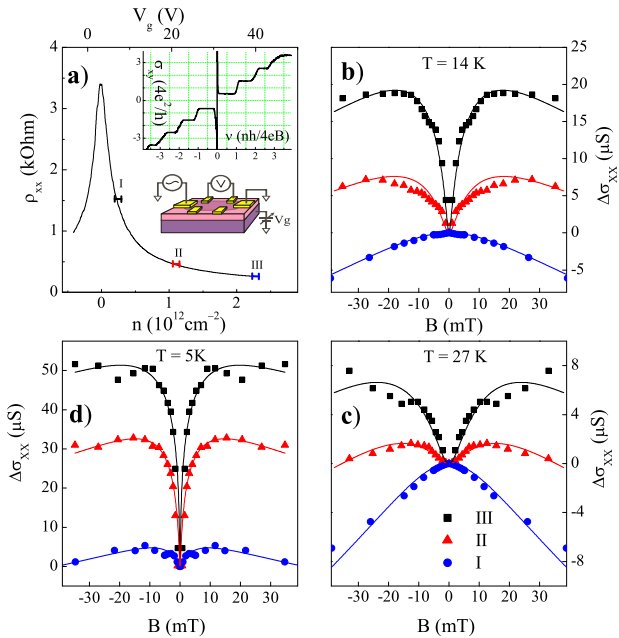}
\caption{(a) Resistivity as a function of the carrier density, with the three regions where the magnetoconductance is studied indicated by bars. Insets: a diagram of the sample and the results of the Quantum Hall effect measurement.
(b,c,d) Evolution of the magnetoconductance with decreasing electron density, at three temperatures. Solid curves are fits to Eq. 1. }\label{fig:two}
\end{figure}

Figures 2b,c,d show the evolution of the MC with changing carrier density at three temperatures: T=5, 14 and 27 K. One can see that with increasing electron density (moving from region III to region I) MC at $T$=14, 27 K changes its sign from positive to negative. Indeed, the ratio of the characteristic times found from the analysis of the MC curves using Eq. 1 is seen to enter the region of anti-localisation as shown in Fig. 1b. Figure 1b also shows that at $T$=5 K, as the value of  $\tau_{\phi}$ is larger, one cannot achieve the transition to anti-localisation because of the large values of $\tau_{\phi}/\tau_*$ and $\tau_{\phi}/\tau_i$.

Figure 3 shows the evolution of the MC with increasing temperature in the three regions of carrier density. It follows from Eq. 1 that at low temperatures the width of the dip in small $B$ is mainly controlled by $\tau_{\phi}$, while the bending of the curve at larger $B$ is determined by $\tau_i$ and $\tau_*$. The analysis of the MC curves shows that, as expected, elastic times $\tau_i$ and $\tau_*$ are essentially temperature independent but inelastic time $\tau_{\phi}$ strongly decreases with increasing $T$. One can see in Fig. 3 that with increasing $T$ the width of the dip in the MC increases (due to a decrease of $\tau_\phi$), so that the dependence becomes flat at $T=27$ K -- the situation where the system does not have a quantum correction to the conductance at $B=0$.  With further increase of the temperature the quantum correction starts being seen again, but now as a peak in the MC. Its width continues to increase with increasing $T$, until at $T\sim$200 K the dependence becomes flat again when anti-localisation disappears due to rapid dephasing of electron trajectories.  Note that the transition from localisation to anti-localisation is seen in regions I and II, but not in the high-density region III.  In this region the inter-valley time $\tau_i$ becomes too small to satisfy the condition of anti-localisation, Fig. 1b.

Our experiments are performed at much higher temperatures than the previous studies of weak localisation: $T<$20 K on single-layer and bilayer exfoliated graphene \cite{Gorbachev, Tikhonenko, Horsell}, and $T<$50 K on few-layer graphene on SiC \cite{Wu} and graphene with antidots \cite{Eroms}. In Fig. 3 one can see that in mechanically exfoliated graphene the temperature dependent MC (negative or positive) exists at temperatures $T\sim200$ K. This is an interesting fact as in conventional 2D systems the quantum correction to the conductance usually disappears at much lower temperatures, due to intensive electron-phonon scattering \cite{Gershenzon}. In graphene, however, electron-phonon scattering is expected to be weak \cite{Guinea,DasSarma}.
\begin{figure}[t]
\includegraphics[width=\columnwidth]{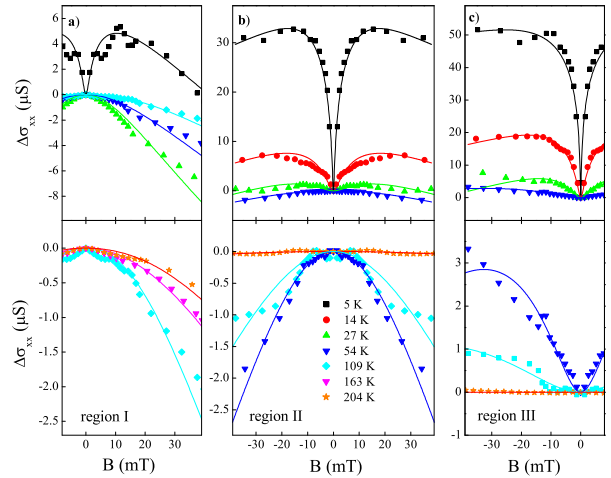}
\caption{(a,b,c) Evolution of the magnetoconductance with increasing temperature in the three studied regions, showing a transition from positive to negative low-field magnetoconductance. Bottom panels are zoomed-in views for high temperatures. Solid curves are fits to Eq. 1. }\label{fig:three}
\end{figure}

In previous studies of electron localisation \cite{Wu, Tikhonenko} the temperature dependence of the dephasing rate $\tau_{\phi}^{-1}$ was found to follow the linear temperature dependence corresponding to electron-electron scattering in the so-called `dirty' or `diffusive' regime \cite{Altshuler}:
\begin{eqnarray}
\tau_{ee}^{-1}&=&\alpha\frac{k_BT}{2E_F\tau_p}\ln\left(\frac{2E_F\tau_p}{\hbar}\right)~,
\end{eqnarray}
where $E_F$ is the Fermi energy, $\tau_p$ is the momentum relaxation time and $\alpha$ is a coefficient of the order of unity. This regime corresponds to the condition $k_BT\tau_p/\hbar<1$, which means that two interacting electrons experience many collisions with impurities during the time of their interaction $\hbar/k_BT$. Figure 4 shows the temperature dependence of the dephasing rate obtained from our analysis of the MC in the three studied regions of carrier density. The expected dephasing rate due to phonon scattering is also shown in the figure, assuming that it is close to the calculated transport phonon scattering rate in graphene \cite{Guinea,DasSarma}:
\begin{eqnarray}
\tau_{ph}^{-1}&=&\frac{1}{\hbar^3}\frac{E_F}{4V_F^2}\frac{D_a^2}{\rho_m V_{ph}^2}k_BT~, \label{eqn:four}
\end{eqnarray}
where $D_a$ is the deformation potential constant, $\rho_m$ is the density of graphene, $V_{ph}$ is the speed of sound, and $V_F$ is the Fermi velocity. (In plotting the electron-phonon rate we have used the values $\rho_m=7.6\times10^{-7}~$kg~m$^{-2}$, $V_{ph}=2\times10^4~$m~s$^{-1}$, $V_F=10^6~$m~s$^{-1}$  and $D_a\approx$ 18 eV which were recently used in the analysis of the temperature dependence of the classical conductance \cite{Fuhrer}.) \begin{figure}[t]
\includegraphics[width=\columnwidth]{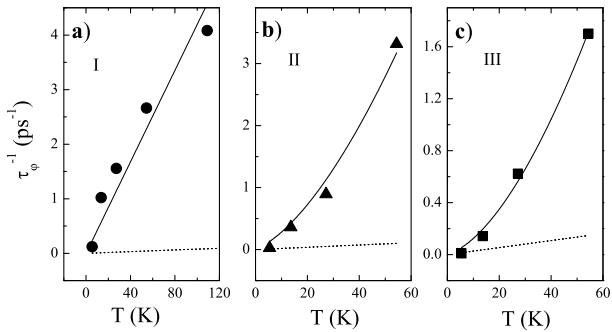}
\caption{Temperature dependence of the dephasing rate for the three regions. Solid curves show fits to the electron-electron scattering rates found as a sum of Eq. 3 and Eq. 5: (a) $\alpha= 1.5, \beta=0$; (b) $\alpha= 1.5, \beta= 2.5$; (c) $\alpha=0, \beta=2.5$. Dotted lines are electron-phonon rates calculated using Eq. 4. }\label{fig:four}
\end{figure}
One can see that the experimental dephasing rate is much higher than the electron-phonon rate and its temperature dependence in regions II and III has a power higher than 1.  This excludes not only electron-phonon but also electron-electron scattering in the `diffusive' regime. The alternative inelastic scattering mechanism is electron-electron scattering in the `ballistic' regime which is realised at $k_BT\tau_p/\hbar>1$, when electrons interact with each other without multiple impurity scattering. It gives a parabolic temperature dependence of the dephasing rate \cite{Narozhny}:
\begin{eqnarray}
\tau_{ee}^{-1}&=&\beta\frac{\pi}{4}\frac{\left(k_BT\right)^2}{\hbar E_F}\ln\left(\frac{2E_F}{k_BT}\right)~.\label{eqn:three}
\end{eqnarray}
This dependence has been observed in earlier experiments on high-mobility 2D systems at low temperatures \cite{Taboryski,Proskuryakov,Price}, with a coefficient $\beta$ of the order of unity. We see that in region I where the transition temperature $k_BT_0\tau_p/\hbar \sim 1$ between diffusive and ballistic regimes is high, $T_0\sim$ 80 K, and the dephasing rate can be satisfactorily explained by the diffusive electron-electron interaction, Eq. 3. In regions II and III with $T_0\sim$60 K and 40 K, respectively, the dephasing rate can be represented as a sum of the two rates, Eq. 3 and Eq. 5. Therefore, our results show that it is electron-electron scattering which is the main source of high-temperature dephasing in graphene.

In summary, we have shown that electron anti-localisation in graphene can be realised experimentally, at small enough values of the dephasing time compared with the elastic inter- and intra-valley scattering times.  This effect is a direct consequence of the unusual nature of the charge carriers: their chirality and the Berry phase $\pi$. We show that quantum interference provides a sensitive tool to detect the presence of defects responsible for inter-valley scattering and chirality breaking of electrons in graphene. We also demonstrate
 that quantum interference in graphene can exist at extremely high temperatures.

We are grateful to D.W.Horsell and I.Gornyi for reading the manuscript and valuable comments, and acknowledge support from the EPSRC grant EP/D031109.


\begin{thebibliography}{99}

\bibitem{Beenakker} C.~W.~J.~Beenakker, and  H.~Van Houten, Solid State Physics \textbf{44}, 1  (edt. by H. Ehrenreich and D. Turnbull, Academic Press Inc., San Diego, 1991).
\bibitem{AKS} A. K. Savchenko, A. S. Rylik, and V. N. Lutskii, Zh. E'ksp. Teor. Fiz. \textbf{85}, 2210 (1983) [JETP 58, 1279 (1983)].
\bibitem{Bergman} G.~Bergman, Phys. Rep. \textbf{107}, 1 (1984).
\bibitem{Meier} C. Schierholz, T. Matsuyama, U. Merkt, and G. Meier, Phys. Stat. Sol. (b) \textbf{233}, 4364 (2002).
\bibitem{Klein} M. I. Katsnelson, K. S. Novoselov  and A. K. Geim, Nature Physics \textbf{2}, 620 (2006).
\bibitem{Kim} A. F. Young, P. Kim, Nature Physics \textbf{5}, 222 (2009).
\bibitem{MorpurgoPRL06} A.~F.~Morpurgo, and F.~Guinea, Phys. Rev. Lett. \textbf{97}, 196804 (2006).
\bibitem{McCannPRL06} E.~McCann \textit{et al.}, Phys. Rev. Lett. \textbf{97}, 146805 (2006).
\bibitem{Morozov} S. V. Morozov \textit{et al.}, Phys. Rev. Lett. \textbf{97}, 016801 (2006).
\bibitem{Wu} X.~Wu, X.~Li, Z.~Song, C.~Berger, and W.~A.~de Heer, Phys. Rev. Lett. \textbf{98}, 136801 (2007).
\bibitem{Gorbachev} R.~V.~Gorbachev, F.~V.~Tikhonenko, A.~S.~Mayorov, D.~W.~Horsell and A.~K.~Savchenko, Phys. Rev. Lett. \textbf{98}, 176805 (2007).
\bibitem{Tikhonenko} F.~V.~Tikhonenko, D.~W.~Horsell, R.~V.~Gorbachev, and A.~K.~Savchenko, Phys. Rev. Lett. \textbf{100}, 056802 (2008).
\bibitem{Horsell} D. W. Horsell, F. V. Tikhonenko, R. V. Gorbachev, and A. K. Savchenko, Phil. Trans. R. Soc. A \textbf{366}, 245 (2008).
\bibitem{NovoselovScience04} K.~S.~Novoselov \textit{et al.}, Science \textbf{306}, 666 (2004).
\bibitem{Monolayer} K.~S.~Novoselov \textit{et al.}, Nature \textbf{438}, 197 (2005).
\bibitem{Kechedzhi} K. Kechedzhi \textit{et al.}, Phys. Rev. Lett. \textbf{102}, 066801 (2009).
\bibitem{Eroms} J. Eroms, D. Weiss, arXiv:0901.0840v1 (2009).
\bibitem{Gershenzon} M.~E.~Gershenson \textit{et al.}, JETP Lett. \textbf{35}, 576 (1982).
\bibitem{Altshuler} Altshuler B. L., Aronov A. G. and Khmelnitsky D. E. J. Phys. C: Solid State Phys. \textbf{15}, 7367 (1982).
\bibitem{Guinea} T.~Stauber, N.~M.~R.~Peres, and F.~Guinea, Phys. Rev. B \textbf{76}, 205423 (2007).
\bibitem{DasSarma} E.~H.~Hwang and S.~Das Sarma, Phys. Rev. B \textbf{77}, 115449 (2008).
\bibitem{Fuhrer} J.~H.~Chen \textit{et al.}, Nature Nanotechnology \textbf{3,} 206 (2008).
\bibitem{Narozhny} B.~N.~Narozhny, G. Zala, and I.~L.~Aleiner, Phys. Rev. B \textbf{65}, 180202 (2002).
\bibitem{Taboryski} R.~Taboryski \textit{et al.}, Semicond. Sci. Technol. \textbf{5}, 933, (1990).
\bibitem{Proskuryakov} Y.~Y.~Proskuryakov \textit{et al.}, Phys. Rev. Lett. \textbf{86}, 4895 (2001).
\bibitem{Price} A.~S.~Price \textit{et al.}, Science, \textbf{316}, 99 (2007).



\end{thebibliography}
\end{document}